\documentclass{appolb}
\usepackage{epsfig}
\begin{document}
\eqsec  % uncomment this line to get equations numbered by (sec.num)
\title{KAON CONDENSATION IN NEUTRON STARS AND HIGH DENSITY
BEHAVIOUR OF NUCLEAR SYMMETRY ENERGY 
\thanks{This research is partially supported by the Polish State Committee 
for Scientific Research (KBN), grants 2 P03B 131 13 and 2 P03D 001 09}%
}
\author{S.  Kubis$^a$ and M. Kutschera$^{a,b}$
\address{$~^a$H.Niewodnicza\'nski Institute of Nuclear Physics\\
Radzikowskiego 152, 31-342 Krak\'ow, Poland\\
$~^b$Institute of Physics, Jagiellonian University\\
Reymonta 4, 30-059 Krak\'ow, Poland}
}
\maketitle
\begin{abstract}
We study the influence of a high density behaviour of the nuclear symmetry
energy on a kaon condensation in neutron
stars. We find that the symmetry energy typical for several
realistic nuclear potentials, which decreases at high densities,
inhibits kaon condensation for weaker kaon-nucleon couplings at
any density.
There exists a threshold  coupling above which the
kaon condensate forms at densities exceeding some critical value.
This is in contrast to the case of rising symmetry energy, as e.g. for 
relativistic mean field models, when
the kaon condensate can form for any coupling at a sufficiently high 
density. Properties of the condensate are also different in both
cases. 
\end{abstract}
\PACS{ 21.65.+f, 97.60.Jd}
  
\section{Introduction}

The possibility  that a charged kaon condensate is present in the ground state
of dense baryon matter  
has been suggested by Kaplan and Nelson [1].
The presence of a kaon condensate  would
strongly affect astrophysically important properties of dense
matter in neutron stars [2]. For example, the proton abundance
of neutron star matter could increase, exceeding the direct URCA
threshold. This would strongly accelerate cooling of neutron stars. Also, the 
formation of metastable neutron stars could be allowed, as the equation of state
of hot matter with
trapped neutrinos would stiffen, supporting larger maximum mass than
the cold one [2]. Metastable neutron stars with masses exceeding the maximum
mass for the cold equation of state would 
collapse to black holes after the neutron star matter
becomes transparent to neutrinos. 

The
formation of a kaon condensate is expected because of the presence 
of strongly attractive interactions between kaons and nucleons, mainly
from the so-called
sigma term, $\Sigma^{KN}$. These
interactions lower the kaon effective mass at higher
densities leading to the condensation of kaons above a
critical density, $n_{kaon}$, which is estimated in Ref.[1] to be 
 $\sim 3 n_0$, where
$n_0=0.16 fm^{-3}$ is the nuclear saturation density.

To study
the possibility of a kaon condensation in neutron stars one should account
for the $\beta$-equilibrium of the neutron star matter which
requires that kaon and electron chemical potentials are equal,
 $\mu_{K^-}=\mu_e$, where
the electron chemical potential is given by a difference of proton and
neutron chemical potentials,  $\mu_e=\mu_P-\mu_N$. This formula shows that also
nucleon-nucleon interactions, which determine both $\mu_P$ and $\mu_N$,
 are of crucial importance for the existence of a
kaon condensation in neutron stars. 
 Thorsson, Prakash and Lattimer [3]
have studied the role of nuclear interactions in 
a kaon-condensed neutron star matter
using simple parametrizations of  nuclear forces.
However,  parametrizations 
used in Ref. [3] are rather similar as far as the corresponding nuclear
symmetry energy, $E_{sym}(n_B)$, is concerned. For all models in Ref.[3], 
$E_{sym}(n_B)$ monotonically increases with increasing baryon density $n_B$.
This  behaviour at high densities
differs considerably from that corresponding to
several realistic phenomenological interactions, such as e.g.
$UV14+TNI$ [4], or $AV14+UVII$ [4], for which the symmetry energy
actually saturates and then decreases at high densities.
In our paper we study consequences of this kind of 
behaviour of the nuclear symmetry energy for the formation of a kaon condensate
in neutron stars.

The electron chemical potential, $\mu_e$, and the proton abundance in the 
neutron star matter are very sensitive to 
the nuclear symmetry energy at high densities [5]. 
Thus, also the critical density for kaon
condensation, $n_{kaon}$, depends on   $E_{sym}(n_B)$.  The
latter quantity is subject to large uncertainties at higher densities [5]. 
As indicated
above, different models which fit the saturation
properties of nuclear matter give rather incompatible predictions of the 
symmetry energy
at high densities. We find that this
uncertainty affects strongly the critical density of the kaon
condensation. In particular, for the symmetry energy which
decreases at high densities, as e.g. for the $UV14+TNI$
interactions, the kaon condensation is inhibited, at least for
weaker kaon-nucleon couplings. This is in contrast to the case of
monotonically increasing symmetry energy when the kaon
condensate can form for any value of the kaon-nucleon coupling at a
sufficiently high density.

In the next section we describe
the model of the kaon 
condensate which includes realistic nucleon-nucleon interactions. In Sect.3 
we discuss the high density behaviour
of the nuclear symmetry energy.  Main results concerning the dependence of
the condensate properties on the symmetry energy are presented in Sect. 4.
\section{The kaon condensate in neutron stars}

The Kaplan and Nelson model [1] of the kaon condensate employs
an $SU(3)\times SU(3)$ 
chiral lagrangian, obtained in the chiral perturbation theory, which involves octets
of pseudoscalar mesons and baryons. 
Brown, Kubodera and Rho [6] have shown that
interactions leading to the kaon
condensation in the Kaplan and Nelson model are dominated by the
s-wave kaon-nucleon coupling. Contributions due to remaining interactions are  
less important and we neglect them here.

When only the s-wave kaon-nucleon
interactions are relevant, the kaon condensate is spatially
uniform and its time dependence is $<K^->=v_K \exp(-i\mu_{K^-} t)$
[7], where $v_K$ is the amplitude of the mean $K^-$-field. 
For such a condensate the effective Kaplan-Nelson lagrangian
in the sector including only nucleons and kaons, that is relevant to the neutron 
star matter, reads [3]:
$$ L_{KN}={f^2  \over 2} \mu^2 sin^2 \theta - 2 m_K^2f^2sin^2
{\theta \over 2} + 
n^{\dag}n(\mu sin^2
{\theta \over 2} -(2a_2+4a_3)m_s sin^2{\theta \over 2}) $$
$$+p^{\dag}p(2\mu sin^2{\theta \over 2} -(2a_1+2a_2+4a_3)m_s sin^2{\theta
\over 2})+L_N. \eqno(1)$$
Here $\mu\equiv \mu_{K^-}$ is the kaon chemical potential,
$\theta=\sqrt{2} v_K/f$ is the "chiral angle", where $f=93 MeV$ is
the pion decay constant, and $L_N$ is the free nucleon lagrangian.  
The parameters, $a_1$, $a_2$, and $a_3$ are coefficients of the interaction terms
in the original Kaplan-Nelson lagrangian which provide splitting of the masses
in the baryon octet [1], and $m_s$ is the strange quark mass. 

 The effective lagrangian (1) contains three effective kaon-nucleon
coupling parameters, $a_1m_s$, $a_2m_s$, and $a_3m_s$ (our notation follows that
of Ref.[3]).  The first two are 
determined by fitting the strange baryon masses [3]. 
In the following we adopt
the values $a_1m_s=-67 MeV$ and $a_2m_s=134 MeV$ from Ref.[3]. The third 
parameter, $a_3m_s$, is related to the kaon-nucleon sigma term,
$$\Sigma^{KN}=-{1 \over 2}(a_1+2a_2+4a_3)m_s, \eqno(2)$$
which is poorly known and thus the value of $a_3m_s$ is  subject to a 
considerable uncertainty. We use here values in the range 
$-134MeV>a_3m_s>-310 MeV$ which correspond to the sigma term in the range 
$170 MeV<\Sigma^{KN}<520 MeV$ [3].
  
The lagrangian (1) describes only the kaon-nucleon interactions in the condensate. 
To account for nucleon-nucleon interactions, which are crucial in a dense nucleon 
matter in neutron stars, we  use realistic models of the nucleon matter [4]. The 
energy density of the neutron star matter with the kaon condensate is composed 
of three contributions,
$$\epsilon_{ns}=\epsilon_{KN}+\epsilon_{N}+\epsilon_{lep},\eqno(3)$$
where $\epsilon_{KN}$ is the energy of the kaon condensate described by 
the lagrangian (1), $\epsilon_{N}$ includes both the nucleon-nucleon 
interaction contribution and the nucleon Fermi energy, and $\epsilon_{lep}$ 
is the electron and muon contribution. 

The energy per particle, $\epsilon_{N}/n_B$, obtained in variational many-body 
calculations
with realistic nucleon-nucleon interactions, 
can be written as a function of the baryon number density, $n_B$, and 
the proton fraction, $x=n_P/n_B$,
in the form [8]
$$E(n_B,x)=T_F(n_B,x)+V_0(n_B)+(1-2x)^2 V_2(n_B), \eqno(4)$$
where $T_F(n_B,x)$ is the Fermi-gas energy, and $V_0(n_B)$ and
$V_2(n_B)$ are the interaction energy contributions. 

The energy density of the neutron star matter with the kaon condensate
including nucleon-nucleon interactions reads
$$\epsilon_{ns}(n_B,x,\mu,\theta) =  T_F(n_B,x)n_B + n_BV_0(n_B) + 
n_B(1-2 x)^2 V_2(n_B)  $$
 $$ + 2 m_K^2 f^2 \sin^2{\theta\over2}+f{\mu^2 \over 2} \sin^2{\theta}   
 + (2 a_1 m_s x + 2 a_2 m_s + 4 a_3 m_s) n_B \sin^2{\theta\over2}  
 + \epsilon_e + \epsilon_\mu ,\eqno(5)  
$$ where the electron Fermi sea contribution is
$$\epsilon_e = {\mu^4\over 4 \pi^2}.  \eqno(6)$$ 
Muons are present only when the electron chemical potential, $\mu$, exceeds the muon rest
mass, $\mu>m_{\mu}$. The energy density of the muon Fermi sea is
$$\epsilon_\mu =  m_\mu^4 f(p_{F_\mu}/m_\mu),\eqno(7)$$ 
with
$$ f(y) = {1\over8 \pi^2}((y + 2 y^3)\sqrt{1 + y^2} - {\rm
arsinh}y).  \eqno(8)$$

The above formulae allow us to determine the critical density 
of kaon condensation and to study properties of the neutron star matter with 
a developed kaon condensate. We obtain the ground state parameters,  
the proton fraction, $x$, the kaon chemical potential, $\mu$,
the kaon condensate amplitude, $\theta$, and the energy density, $\epsilon_{ns}$,
as functions of the baryon number density, $n_B$, by
optimization of the
thermodynamical potential 
$\tilde{\epsilon}  = \epsilon_{ns} - \mu (n_K^-+n_e+n_\mu-n_P)$ 
which reads
$$ 
\tilde{\epsilon}(n_B,x,\mu,\theta) =   T_F(n_B,x)n_B + 
n_BV_0(n_B) +  n_B (1-2 x)^2 V_2(n_B)   $$
$$   -f^2 {\mu^2\over2} \sin^2 \theta  + 2 m_K^2 f^2 \sin^2{\theta\over2}   
     +\mu n_B x   -  \mu n_B (1+x) \sin^2{\theta\over2} $$
$$   +(2 a_1 m_s x + 2 a_2 m_s + 4 a_3 m_s) n_B \sin^2{\theta\over2}  
     +\tilde{\epsilon}_e + \tilde{\epsilon}_\mu. \eqno(9)
$$
This thermodynamical potential
is related to the (Landau and Lifshitz) potential, 
$\Omega/V = \epsilon - \Sigma \mu_i n_i = \tilde{\epsilon} - \mu_N n_B$,
where $\mu_N$ is the neutron chemical potential.

The critical density, 
$n_{kaon}$, is defined as a density  at which the condensate amplitude starts to
deviate from zero. Our procedure provides the chiral angle of the condensate
as a function of density, $\theta(n_B)$. The value of $n_{kaon}$ is thus found 
from the condition $\theta(n_{kaon})=0$. 

\section{ The nuclear symmetry energy at high densities}

The nucleon-nucleon interaction energy is parametrized in Eq.(4) in terms of 
the isoscalar and isovector contributions, $V_0(n_B)$ and $V_2(n_B)$.
As one can notice, at a given baryon density, $n_B$, 
the interaction energy density corresponding to the isoscalar part, 
$n_BV_0(n_B)$ in Eq.(9), is 
constant and thus it does not
play any role in the optimization of the potential $\tilde\epsilon$ which 
determines the kaon condensate parameters. It is the isovector contribution, 
$V_2(n_B)$, which is 
crucial for the onset of the condensation. This component is directly
related to the nuclear symmetry energy which expressed in terms of $V_2(n_B)$ 
reads
$$ E_{sym}(n_B)={5 \over 9} T_F(n_B, {1 \over 2})+V_2(n_B). \eqno(10)$$

As we already mentioned,
the high density behaviour of $E_{sym}(n_B)$ is not well known
at present. Different model calculations give
incompatible extrapolations away from the empirically
determined value at $n_0$, $E_{sym}(n_0)=31 \pm 4 MeV$ [9].
In some approaches [5],
$E_{sym}(n_B)$ is a monotonically increasing function of the baryon
number density, whereas other models [5] predict $E_{sym}(n_B)$ to
saturate and then decrease at high densities. Physically, in the
former case the energy of pure neutron matter is always higher
than the energy of symmetric nuclear matter, while in the latter
one pure neutron matter becomes eventually the ground state of
dense baryon matter.
This difference has profound consequences for neutron star
matter. Here we study how it influences the onset of the kaon
condensation. 
\begin {figure}[h]
\epsffile {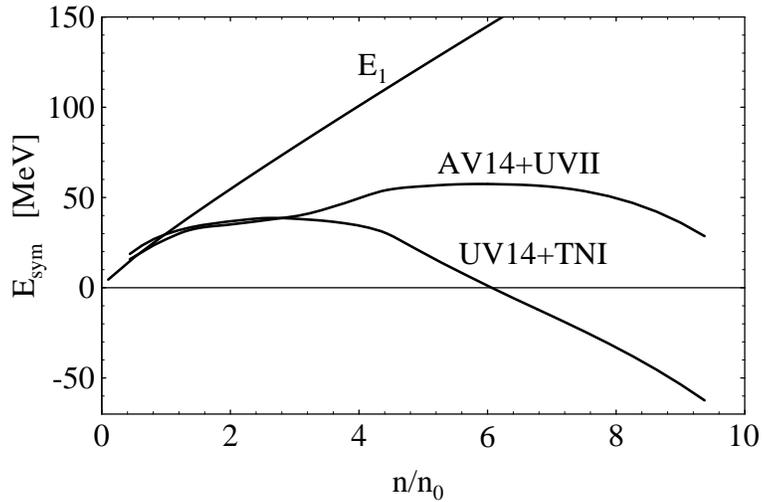}
\caption{ The nuclear symmetry energy as a function of density for 
indicated models of nuclear interactions. The curve $E_1$ corresponds to a 
linear parametrization of Ref.[3].}
\end{figure}

In Fig.1 we show the symmetry energy corresponding to three different models
of nucleon matter which we use in our calculations. The curve labelled $E_1$ 
corresponds to a linear parametrization used in Ref.[3]. It fits the empirical value, 
$E_{sym}(n_0)$, and then extrapolates linearly to higher densities. Such a linear
dependence  of the symmetry energy on density is typical for
relativistic mean field models of nucleon matter [5]. Other curves in Fig.1 show
the symmetry energy from Ref.[4] obtained in variational many-body calculations 
with realistic two-body nucleon-nucleon potentials and three-body interactions.
The curves labelled $UV14+TNI$ and $AV14+UVII$ correspond to the Urbana 
$v_{14}$ two-body potential with the
TNI three-body term [4]
and the Argonne $v_{14}$ potential with the UVII three-body interaction [4], 
respectively. 
As one can notice, realistic models predict that the symmetry energy saturates
and then decreases with increasing baryon density. This behaviour is markedly
different from that of the curve $E_1$. 

\section { The critical density and properties of the kaon-condensed neutron star 
matter}

In this section we present results for three models of the nuclear symmetry 
energy shown in Fig.1. Particularly interesting is the corresponding critical
density, $n_{kaon}$. To best illustrate the influence of the symmetry energy on
$n_{kaon}$ we use a different definition of the critical density which states
that $n_{kaon}$ is the lowest density for which the
energy  of the lowest state of $K^-$ in the neutron star matter, $\omega_-$,
becomes equal to the electron chemical potential,
$$ \omega_-=\mu. \eqno(11)$$
At higher densities, $n>n_{kaon}$, kaons form a Bose-Einstein condensate of 
finite amplitude. This definition of the critical density is equivalent to that
given at the end of Sect.2.

 The lowest kaon energy in the nucleon medium reads:
$$ \omega_-=-n_B(1+x) {1 \over 4f^2}+$$ 
$$+\Bigl( {1 \over
16f^4}n_B^2(1+x)^2+m_K^2+{n_B \over 2f^2}(2a_1x+2a_2+4a_3)m_s\Bigr)^{1 \over 2}. \eqno(12)$$
\begin{figure}[h]
\epsffile {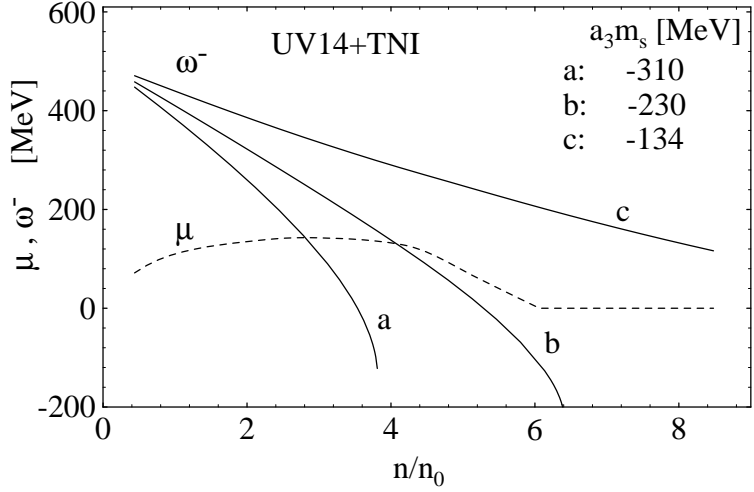}
\caption{The lowest kaon energy in the nucleon matter, $\omega_-$, (solid curves) and
the electron chemical potential, $\mu$, (dashed curve) for the 
UV14+TNI interactions. Curves labelled a,b, and c 
correspond to the coupling parameter $a_3m_s$ of $-310 MeV$, $-230 MeV$, and
$-134 MeV$, respectively.}
\end{figure}
The energy  $\omega_-$ decreases with
baryon number density, Figs. 2 and 3. 
The condition (11) can be satisfied only at a sufficiently high
density provided the electron chemical potential does not
decrease too fast. In Fig.2 we show the lowest kaon energy, $\omega_-$, for three values 
of the coupling parameter $a_3m_s$, and for the electron
chemical potential, $\mu$, corresponding to the $UV14+TNI$ interactions.
The electron chemical potential is rather low, it has a maximum at $n=3n_0$ and
then decreases to zero at a density of about $6n_0$, where electron and proton 
densities vanish.  The critical condition (11) can be satisfied only
for the coupling values at least as strong as in case (b), $a_3m_s\le -230 MeV$.
For weaker coupling the $\omega_-$ curve does not cross the chemical potential
and the critical condition (11) can never be satisfied. Hence, in the case of the 
symmetry energy 
of the $UV14+TNI$ potential there exists a threshold value of the 
kaon-nucleon coupling parameter, $ a_3m_s $, below which there exists no 
kaon condensation in the neutron star matter.
\begin{figure}[h]
\epsffile {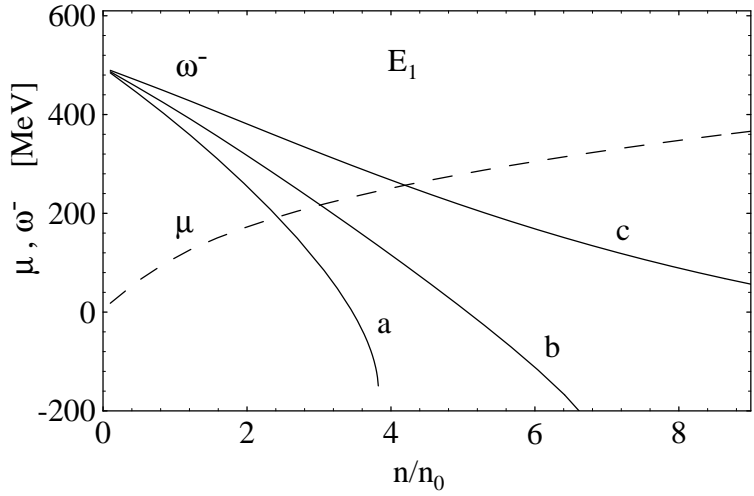}
\caption{The same as in Fig.2, for the $E_1$ symmetry energy}
\end{figure}

Such a threshold does not exist in Fig.3 where we show the kaon energy 
$\omega_-$ and the electron chemical potential for the symmetry energy $E_1$. 
Since the electron chemical potential monotonically increases with density, the
critical condition (11) can be satisfied for any value of the coupling parameter
$a_3m_s$ at a sufficiently high density. Critical densities corresponding to
indicated values of $a_3m_s$ in Fig.2 differ by a factor of about two.

Values of the critical density can also be read off from Fig.4 and Fig.5, where
we show the 
amplitude of the condensate, $\theta$, as a function of density,
for the $UV14+TNI$
and $AV14+UVII$ interactions, respectively. In both figures, results corresponding
to the symmetry energy $E_1$ are also shown. The values of $n_{kaon}$ found from
Fig.2 and Fig.3 coincide with those from Fig.4. It is also interesting to compare how the
angle $\theta$ varies with density.
In Fig.4, for the $UV14+TNI$ model
 the amplitude grows very
fast with density for densities above the critical one. It reaches a maximum
when the proton fraction becomes $x=1$, and decreases at higher densities.
A similar behaviour  is  displayed in Fig.5 for the $AV14+UVII$
interactions for the strongest coupling (solid curve a). This is in contrast to
the $E_1$ case, when the condensate amplitude monotonically grows with density
for all indicated values of the coupling parameters.

\begin{figure}[h]
\epsffile{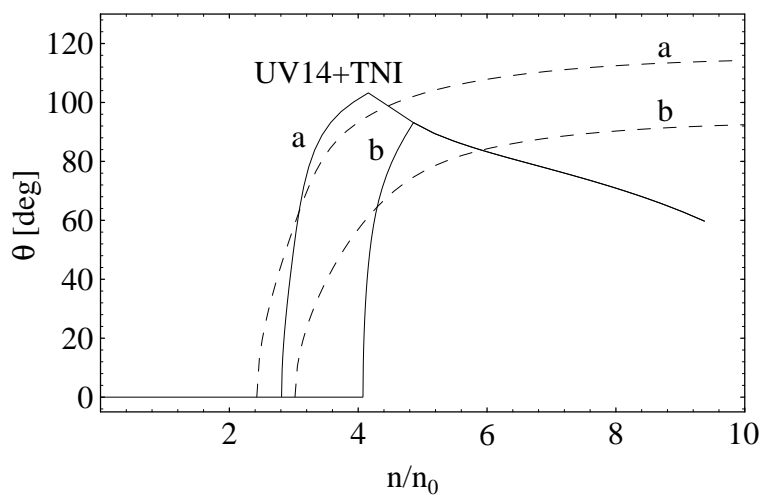}
\caption{The condensate amplitude as a function of density for the $UV14+TNI$
interactions, solid curves, and for the symmetry energy $E_1$, dashed curves. 
The coupling parameters
corresponding to labels a and b are the same as in Fig.2}
\end{figure} 

\begin{figure}[h]
\epsffile{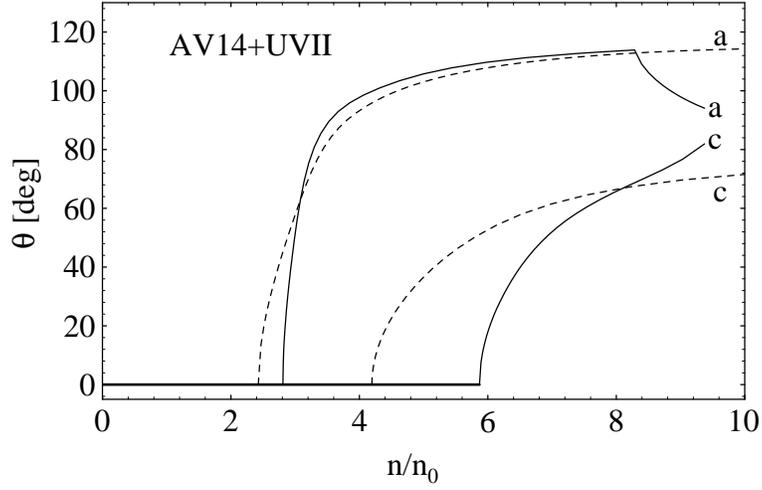}
\caption{The same as in Fig.4 for the $AV14+UVII$ interactions.}
\end{figure}

In Fig.6 the proton fraction of the neutron star matter with the kaon condensate
is shown. One can notice that for the $UV14+TNI$ interactions the kaon-condensed
phase contains only protons. Neutrons are fully converted into protons at 
densities
only slightly exceeding the critical value. This behaviour is similar for all
coupling parameters stronger than the threshold value 
$a_3m_s=-230 MeV$. In Fig.7, where the same is shown for the $AV14+UVII$ 
interaction, one can notice that for the strongest coupling the proton fraction
also reaches unity at high densities.
For interactions with the symmetry energy $E_1$ the proton fraction at high 
densities tends to an asymptotic value $x=0.6$, for all kaon-nucleon couplings.

\begin{figure}[h]
\epsffile{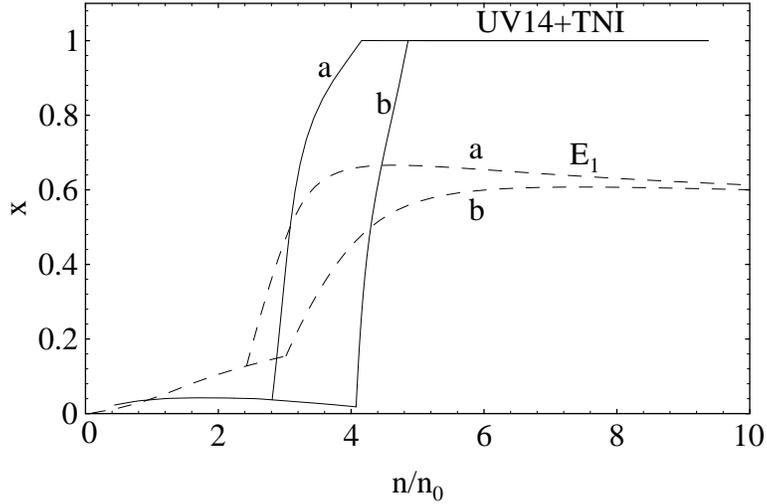}
\caption{The proton fraction of the neutron star matter with a kaon condensate
for the $UV14+TNI$ interactions and for the symmetry energy $E_1$. The curves
a and b correspond to values of the coupling parameter given in Fig.2}
\end{figure}

\section {Discussion}

The above results show that the critical density of kaon condensation in 
neutron stars is rather sensitive to the high density behaviour of the nuclear
symmetry energy. For some realistic equations of state, such as the 
$UV14+TNI$ one, the existence of the condensate is even forbidden if the 
kaon-nucleon interaction is not strong enough. Also, properties of the neutron
star matter with developed condensate are sensitive to the symmetry energy. One can 
notice in Fig.4
and Fig.5 that the condensate amplitude, measured by the angle $\theta$, 
behaves  in a quite different way for considered  models of nucleon-nucleon 
interactions. In particular, the kaon condensate weakens at high densities for
the $UV14+TNI$ interactions, Fig.4. 

For astrophysical applications, the proton
fraction of the neutron star matter plays an important role. The influence
of the symmetry energy on the proton abundance in the kaon-condensed phase is
rather strong, as shown in Fig.6 and Fig.7. In fact, the proton abundance at
higher densities is determined entirely by the nuclear symmetry energy. The
kaon-nucleon coupling parameter, $a_3m_s$, affects the proton fraction only
at densities close to the critical value. The role of the symmetry energy is 
most spectacular for
the $UV14+TNI$ interactions for which neutrons are fully converted into protons,
Fig.6. 

\begin{figure}[h]
\epsffile{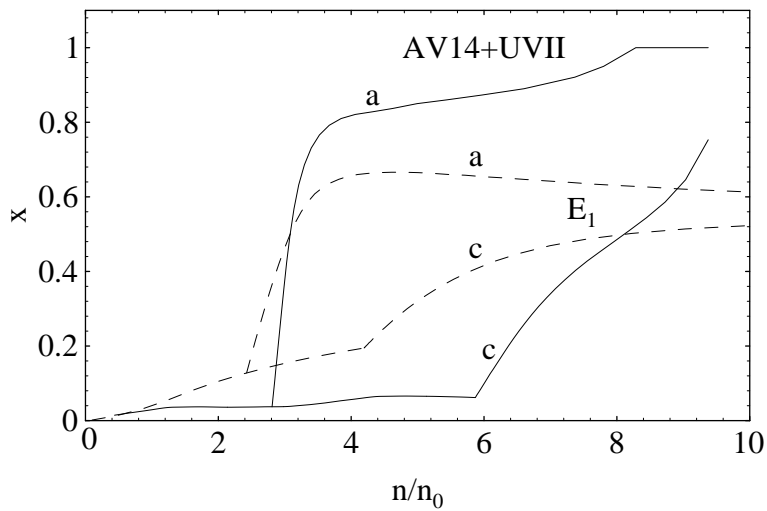}
\caption{The same as in Fig.4 for the $AV14+UVII$ interactions. The curves a and
c correspond to values of the coupling parameter given in Fig.2}
\end{figure}

Let us remark finally that recent claims [10] that all modern phase
equivalent potentials which fit accurately the $n-n$ and $n-p$
scattering data yield a symmetry energy which increases with
density, are unjustified. For the Argonne potential $AV18$ authors
of Ref.[10] find increasing $E_{sym}(n_B)$. Calculations reported in Ref.[11]
show that the symmetry energy corresponding to this
potential saturates and then decreases at high densities in a
similar way as found by Wiringa, Fiks and Fabrocini [4]  for
Urbana $UV14$ and Argonne $AV14$ potentials. Premature
conclusions of Ref.[10] stem from the use of the lowest order
Brueckner approach which is inadequate at high densities.
Hence the uncertainty as to the high density behaviour of the
nuclear symmetry energy is still present. For astrophysical
applications, both decreasing and increasing symmetry energy
should be used in order to assess the role of this
uncertainty.

\end{document}